\documentclass[final,1p,times]{elsarticle}
\usepackage{subcaption}
\usepackage[switch]{lineno}

\begin{document}

\begin{frontmatter}

  \title{Pseudorapidity dependence of anisotropic flow in heavy ion collisions with ALICE}
  \author{Freja Thoresen (on behalf of the ALICE Collaboration)}
  \address{Niels Bohr Institute, Blegdamsvej 17, 2100 Copenhagen, Denmark}

\begin{abstract}
  \begin{small}
    The pseudorapidity dependence of flow coefficients, $v_n$, is presented over a large range in pseudorapidity ($-3.5 \leq \eta \leq 5$) in Pb$-$Pb collisions at 5.02 TeV.
    The flow coefficients are measured with 2- and 4-particle cumulants with $|\Delta \eta | > 0$ and $|\Delta \eta | > 2$. The difference between $v_2\{2,|\Delta \eta|>0\}$ and $v_2\{2,|\Delta \eta|>2\}$ implies changes in either the de-correlation of flow vectors or non-flow effects between the two methods. The AMPT model shows a qualitative agreement with the measurements.
 \end{small}
\end{abstract}

\end{frontmatter}

\section{Introduction}
At the LHC, a hot and dense medium is created in Pb$-$Pb collisions. It is commonly referred to as the Quark-Gluon-Plasma (QGP).
In heavy-ion collisions, the rapid expansion of the system causes the particles to be boosted by a common velocity field. Spatial anisotropies arise from the overlap of the two heavy ions or fluctuations in the initial state. The spatial anisotropies cause anisotropy in the momentum space of the created particles through the evolution of the system.
The azimuthal flow can be described by the coefficients $v_n$ of a Fourier series decomposition of emitted particles \cite{Voloshin:1994mz},
\begin{equation}
  \frac{\mathrm{d}N}{\mathrm{d} \varphi} \propto \frac{1}{2 \pi} [1 + 2 \sum^\infty_{n=1} v_n \cos (n[\varphi - \Psi_n])]
\end{equation}
with
\begin{equation}
  v_n = \langle \cos (n [\varphi - \Psi_n]) \rangle.
\end{equation}

The pseudorapidity dependence of the flow coefficients can give information about the shear viscosity ($\eta/s$) of the system, e.g., temperature dependence of $\eta/s$ \cite{Okamoto:2017rup}. At forward rapidities, there is a shorter lifetime in the QGP phase, which can lead to hadronic viscosity to play a more significant role.
In \cite{Aad:2014eoa, Sirunyan:2017igb, ATLAS:2012at} measurements of $v_2$ was presented in the central regions of pseudorapidity for Pb-Pb at 5.02 and 2.76 TeV at the LHC. Measurements of $v_n$ were extended to forward rapidities in Pb$-$Pb collisions at 2.76 TeV \cite{Adam:2016ows} and in Au$-$Au collisions at 130 GeV \cite{Adams:2005dq}.

\section{Experimental Setup}
The data used in the analysis is collected with the ALICE experiment. The Time Projection Chamber (TPC) is used for tracking of particles in the pseudorapidity region $-1.1 < \eta <1.1 $. The Inner Tracking System (ITS) is used in combination with the TPC for tracking and is also used for triggering. At large pseudorapidities the Forward Multiplicity Detector (FMD) is placed in both the forward  ($1.7 < \eta < 5.0$) and backward direction ($-3.5 < \eta < -1.7$). The VZERO scintillators are also placed at large rapidities ($-3.7 < \eta < -1.7$ and $2.8 < \eta < 5.1$) and are used for triggering and determination of centrality.

\section{Methods}
\begin{figure}[h]
  \centering
  \begin{subfigure}[t]{0.45\textwidth}
    \includegraphics[width=1.0\linewidth]{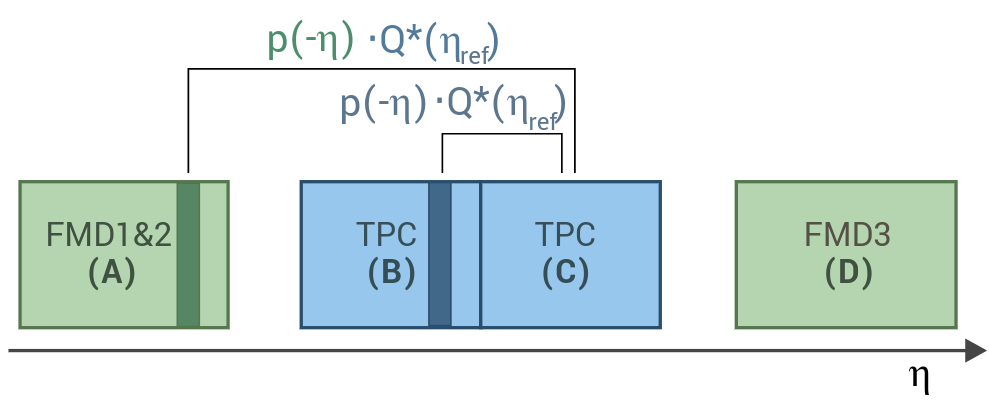}
    \caption{TPC as reference region.}
  \end{subfigure}
  ~~~
  \begin{subfigure}[t]{0.45\textwidth}
    \includegraphics[width=1.0\linewidth]{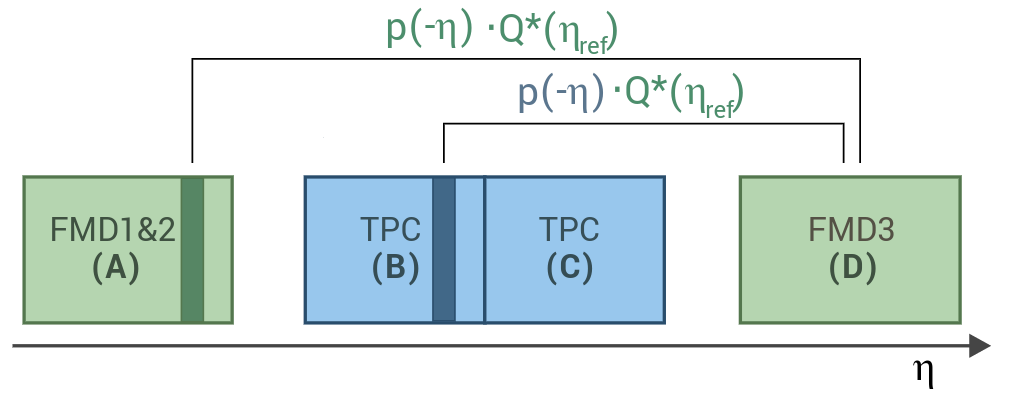}
    \caption{FMD as reference region.}
  \end{subfigure}
  \caption{The two methods used for calculating $v_n(\eta)$. $p$ is the differential vector and $Q$ is the reference vector.}
  \label{fig:methods}
\end{figure}

The flow coefficients $v_n$ are calculated using the Generic Framework \cite{Bilandzic:2013kga}.
In the analysis two methods with sub-events were used. Using the TPC as the reference region, means that there is $|\Delta \eta| > 0$ in the TPC and $|\Delta \eta| > 2$ in the FMD (Figure \ref{fig:methods} left).
If there are no de-correlation of the symmetry planes as a function of $\eta$ and no breaking of factorization, then the calculation for differential flow $v_n'$ in the sub-event $B$ is

\begin{equation}
  v_n^{'B} = \frac{\langle v_n^{'B} v_n^C \rangle }{\sqrt{\langle v_n^B v_n^C \rangle } },
\end{equation}
where $v_n^B$ and $v_n^C$ is reference flow in region $B$ and $C$. The reference region is always chosen to be opposite in $\eta$ to the differential region. E.g. particles of interest in region $A$ are correlated with the reference particles from region $C$.

If one use the FMD as reference region instead, there is $\Delta \eta > 2$ in the TPC and $\Delta \eta > 4$ in the FMD (Figure \ref{fig:methods} right). In this case for e.g. $v_n'$ in region $B$ the calculation is,

\begin{equation}
  v_n^{'B} = \frac{\langle v_n^{'B} v_n^D \rangle }{\sqrt{\langle v_n^A v_n^D \rangle } }
\end{equation}

\section{Results}

\begin{figure}
  \centering
    \includegraphics[width=0.9\linewidth]{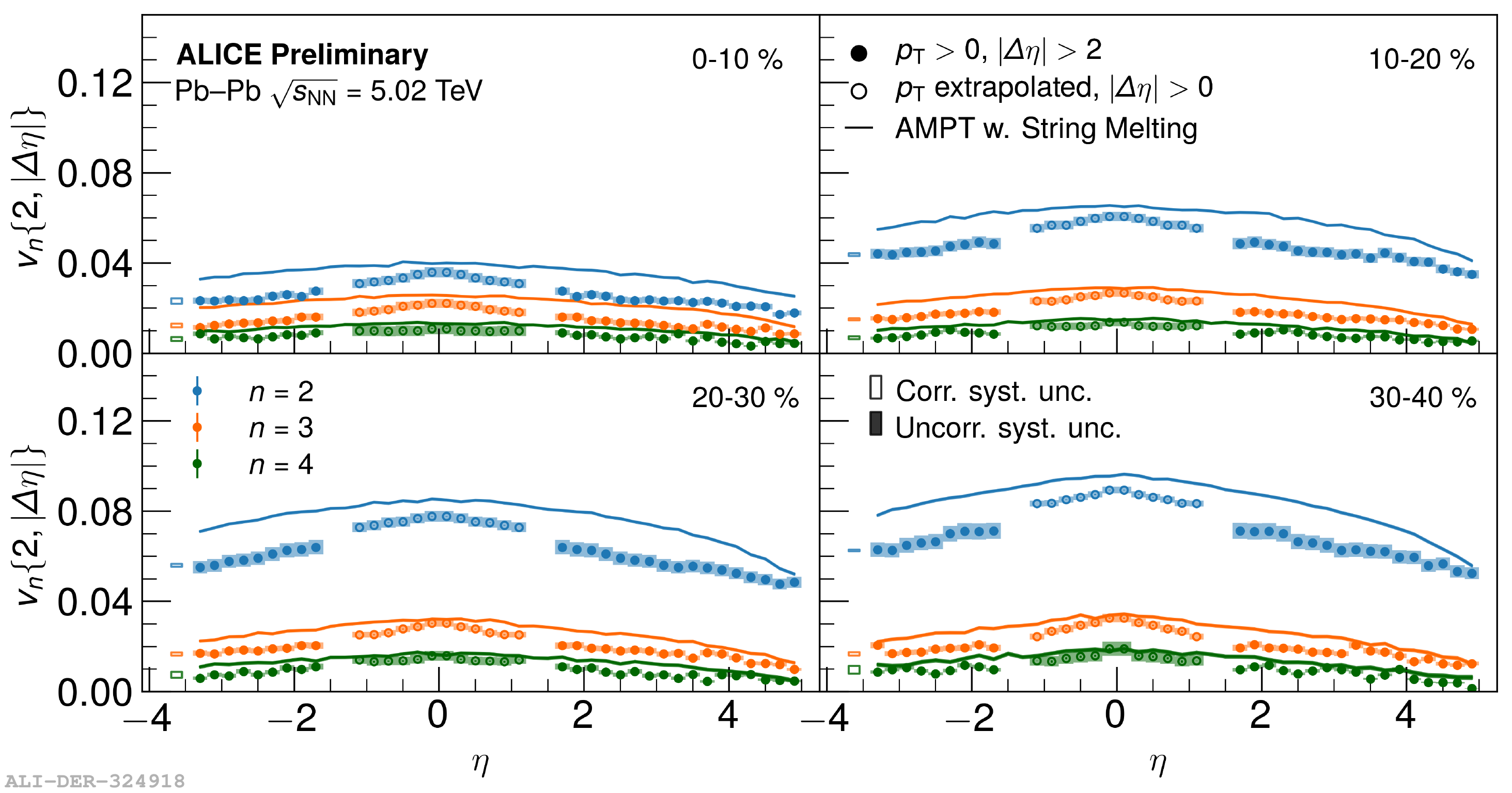}
  \caption{Measurements of $v_n\{2\}$ using the TPC as reference region. The boxes are the systematical uncertainties (divided into correlated and uncorrelated uncertainties) and the statistical uncertainties are mostly smaller than the marker size.}
  \label{fig:vn2}
\end{figure}

Figure \ref{fig:vn2} shows the two-particle cumulant results with TPC as reference region. There is a clear ordering of the flow coefficients, $v_2 > v_3 > v_4$ and $v_n$ is increasing from central to semi-central collisions. $v_3$ and $v_4$ have a much weaker centrality dependence than $v_2$, which indicates that the higher orders are mostly driven by initial state fluctuations.
The $v_n$ measurements are overlayed with simulation results from AMPT with string melting \cite{Lin:2004en}. The AMPT simulation describes the data qualitatively but not quantitatively.

\begin{figure}
  \centering
    \includegraphics[width=0.9\linewidth]{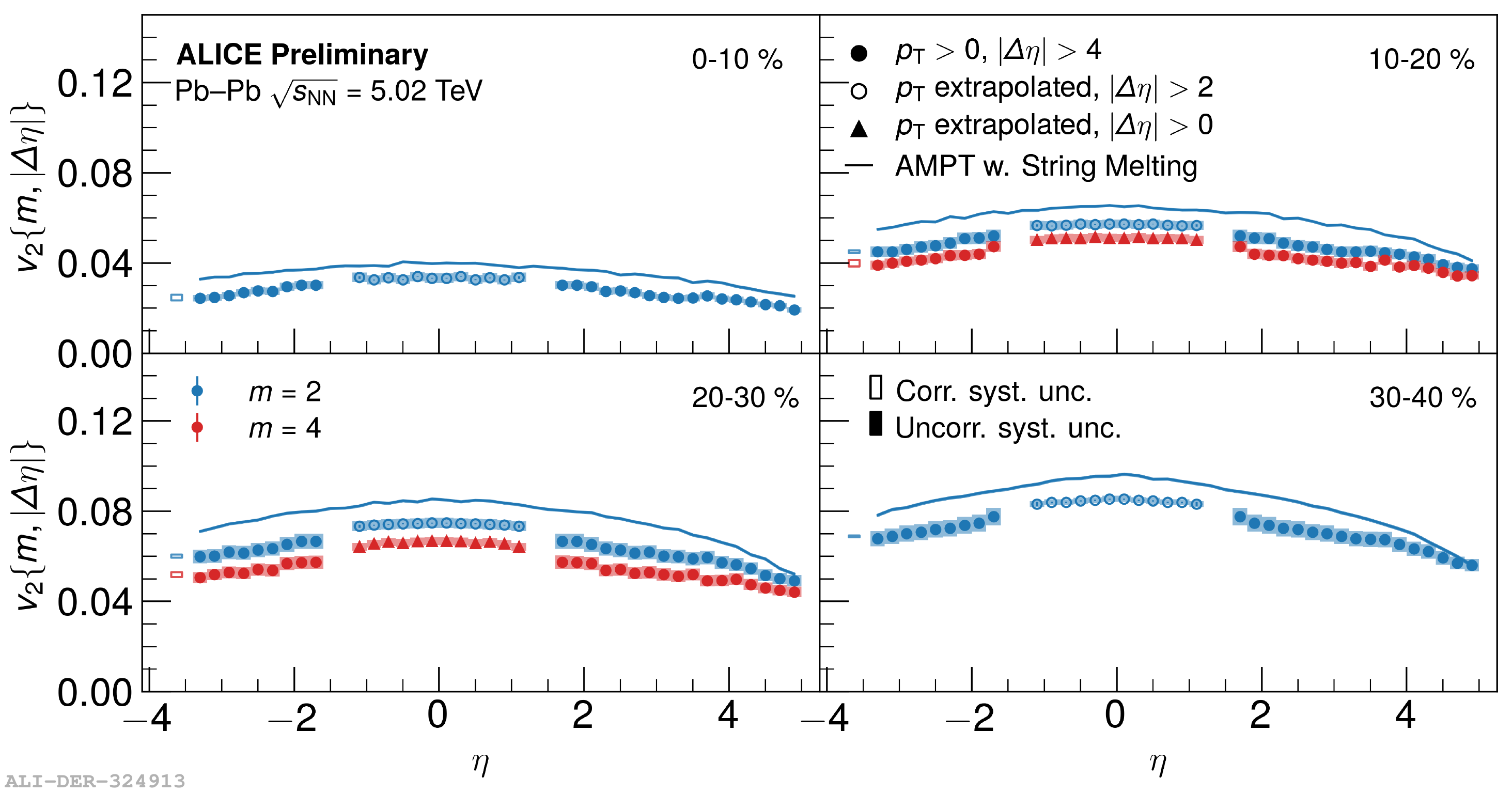}
  \caption{Measurements of $v_2\{m\}$ using 2-particle cumulants with a large $\eta$-gap and $4$-particle cumulant. The boxes are the systematical uncertainties (divided into correlated and uncorrelated uncertainties) and the statistical uncertainties are mostly smaller than the marker size.  }
  \label{fig:v22}
\end{figure}

In Figure \ref{fig:v22}, the results with a large $\eta$-gap with the FMD as reference region are shown, and the 4-particle cumulant results.
The difference in the magnitude between $v_2\{2\}$ and $v_2\{4\}$ is from flow fluctuations. The similar shape of $v_2\{2\}$ and $v_2\{4\}$ could imply that $v_2\{2\}$ is without non-flow, since 4-particle cumulants are known to suppress non-flow effects.
 In the middle pseudorapidity region, there is a clear difference compared to Figure \ref{fig:vn2} by the flat shape. The flat shape implies that there are some effects that a small $\eta$-gap might not be able to suppress, e.g., non-flow and factorization effects.  The AMPT model, also presented for comparison, again, does not describe the data quantitatively.

\section{Summary}
In this proceeding, we present the pseudorapidity dependence of flow coefficients covering a wide range in pseudorapidity in Pb$-$Pb collisions at 5.02 TeV.
The results show that $v_n$ measured with a large $\eta$-gap and 4-particle cumulant has a flatter shape than the 2-particle cumulant with a small $\eta$-gap. The change of the shape of $v_2\{2\}$ when using either long-range or short-range correlations implies changes in either de-correlation of flow vectors or non-flow effects between the two methods. The measurements of $v_3$ and $v_4$ do not have a strong centrality dependence and are therefore mostly driven by fluctuations.
AMPT has a qualitative agreement with presented measurements, but further improvements are needed. In the future, 3+1D hydrodynamic calculations might be compared to constrain initial state models and study the longitudinal dynamics of the hot and dense matter.

\clearpage

\bibliographystyle{unsrt}
\bibliography{references}

\end{document}